\documentstyle[12pt,epsf]{article}
\newcommand{\ra}{\rightarrow}

\author{P.\ M. Nadolsky\\
Institute for High Energy Physics\\
Protvino, Moscow region,\\142284 RUSSIA}
\title{\bf Probing polarized valence quark distributions in
$\bf W^{\bf -}$-boson production}
\date{March 22, 1995}
\begin{document}
\maketitle
\begin{abstract}
While EMC-SMC-E142 data sheds new light on the behavior
of polarized parton distributions at small and intermediate $x$, it
may be also interesting to study valence quark polarizations at large $x$,
where spectator quark counting rules and Carlitz-Kaur type models
predict different behavior for down quark polarizations. The article examines
the possibility of testing the ratio $\Delta d_V(x)/d_V(x)$ at $x\to 1$ in
the inclusive production of $W^{-}$-bosons.
\end{abstract}

In the last years much
attention of HEP community is attracted to the physics of polarized
hadron interactions, to great extent due to the recent interesting data on
polarized inclusive~\cite{EMCSMCE142} and semi-inclusive~\cite{Wislicki}
deep-inelastic scattering.  The EMC-SMC-E142
data already allow to make several conclusions about the nucleon spin
structure at $Q^2\simeq 2\div 10\; GeV^2~$ and values of
Bjorken $x\simeq 0.003\div 0.7$. The analysis of these data in the framework
of perturbative QCD provides information on longitudinal polarized parton
distributions $\Delta q_i(x,Q^2)$, interpreted as the differences of
probabilities $q^{+/-}(x,Q^2)$ for finding partons of the type $i$ with spin
parallel/antiparallel to the spin of the parent nucleon.
 However in the region of  $x\geq 0.1$ the experimental
errors have a tendency to grow, while at $x\geq 0.7$
there is still no data at all.
On the other hand, precise studies of various polarized processes
at large $x$ may provide valuable information about the
behavior of longitudinal valence quark polarizations $\Delta q_V(x)$, that
are dominant in this region compared to sea quarks and gluons.

So far the common opinion on how the polarized quark distributions
$\Delta u_V(x)$ and $\Delta d_V(x)$ should behave at large $x$ is not
fully established.
The experimental information in this region
is not precise enough to  fix $\Delta q_V$'s unambiguously. The
theoretical description is also not unique. This problem was
examined as far as in mid-70's, but not solved to the end.
Since that time the majority of
theoretical  models, describing valence quarks at $x\sim 1$, falls into two
general categories.

The
qualitative analysis of the distinctions between these two approaches can be
based on the representation of a nucleon as composed from the valence quark
carrying a large portion of nucleon's momentum, and a diquark
with $x\sim 0$~\cite{Close};
in good approximation the contribution of the sea may be neglected.  The
experimental results (for example, the behavior of the ratio of nucleon
structure functions $F^n_2(x,Q^2)/F^p_2(x,Q^2)$ at $x\to 1$) convincingly
indicate that the $SU(6)$-symmetry of the nucleon wavefunction is violated,
so that the states containing diquarks with certain spin-isospin numbers
$S,\, S_z,\, I,\, I_z$ are suppressed compared to the situation of the exact
$SU(6)$-symmetry.

The models of the first type follow the works of Carlitz and
Kaur~\cite{CK} and assume that diquarks in this region must be in a $S=0$
rather than in a $S=1$ state.  On the other hand, the models motivated by
perturbative QCD start from the assumption that only
diquarks in a $S_z=1$ state are suppressed. One
of the pioneering works in the framework of the second approach was
published by Farrar and Jackson~\cite{FJ}.

The different mechanism of $SU(6)$-breaking in these two approaches leads
to \quad different \quad predictions \quad for \quad several \quad
quantities
measured \quad both in unpolarized and polarized reactions. The most known of
them is the ratio $F^n_2(x,Q^2)/F^p_2(x,Q^2)$, which tends in these models
correspondingly to $1/4$ or $3/7$. Most part of the existing data on the
structure functions $F^{p,n}_2$ at large $x$ evidences in favor of $1/4$
value, providing that perturbative-QCD result is not valid. As for the
polarized nucleon interactions, so far no experiment, allowing
to discriminate one model from another, was staged on the existing colliders.

Such an experiment can be based on the fact that Carlitz-Kaur (CK) and
Farrar-Jackson (FJ) models disagree about the behavior of valence $d$-quark
polarizations at large $x$. While in CK model
$\Delta d_V(x)/ d_V(x)\to -1/3$, the FJ approach predicts that $\Delta
d_V(x)/ d_V(x)\to 1$.  The information about the limiting behavior of $\Delta
d_V(x)$ may be obtained in various experiments, for instance,
deep-inelastic scattering;
however, it may also be interesting to
carry out a straightforward measurement of the ratio $\Delta d_V(x)/d_V(x)$.

If at least one
polarized proton beam is available, one can measure longitudinal single-spin
asymmetries $A_L$ in the inclusive production of $W^-$-bosons $p^\ra p\to W^-
X$ or $p^\ra \bar p\to W^- X$; in the appropriate kinematical range these
asymmetries will be directly proportional to the ratio $\Delta d_V(x)/
d_V(x)$~\cite{Soffer}.
In fact, it is not necessary to measure $W^{-}$-asymmetries in the region of
$x$ very close to 1. The behavior of $\Delta d_V(x)$ in CK and FJ models
should be very different already at $x\sim 0.5\div 0.7$, where the
cross-sections are still noticeable. This paper is dedicated to the
discussion of the opportunity to study polarized valence distributions in
$W^{-}$-boson production, especially to the explicit evaluation of the
kinematical region, where the event rates will be high enough to make
definite conclusions about the magnitude of $\Delta d_V(x)/d_V(x)$. But, to
begin, let us briefly review the description of valence quark
distributions in CK and FJ models.

Since the behavior of parton distributions is determined
by the  properties of hadrons at large distances, where the running
coupling of quark-gluon interactions is not small, the perturbative QCD
arguments are not directly relevant to their analysis at arbitrary
values of Bjorken $x$. However, the situation when the struck parton
carries almost all of the nucleon's momentum corresponds to very far off-shell
configuration of nucleon constituents~\cite{GunBrod,FJ,BBS}.
Such a configuration can be obtained from the nucleon state with the lowest
orbital momentum and the finite invariant
masses of partons only by exchange of hard gluons. In this case the
dominant contribution to the amplitude of deep-inelastic scattering is
provided by Feynman graphs with minimal number of gluon propagators tying
quarks into a single hadron.

The straightforward calculation of these graphs
shows that the falloff of helicity-dependent quark distributions
is described by the power-law dependence,
\begin{equation}
  q^{+/-}(x)\sim (1-x)^p,
\end{equation}
where
\begin{equation}
 p=2n - 1 +2 \Delta S_z,  \label{power}
\end{equation}
$n$ is the minimal number of spectator quarks and
$\Delta S_z= 0$ for $q^{+}(x)$ or 1 for $q^{-}(x)$. The slowest falloff
corresponds to the valence quark distributions, for which $p=3$.
The essential feature of
(\ref{power}) is that the power-law falloff does not depend on the
flavor of quarks. Therefore the perturbative result prescribes
the distributions of valence up and down quarks to be proportional in
the limit $x\to 1$, and the coefficient of proportionality can be found to be
$1/5$~\cite{BBS,PN}.
It also implies that, independently of quark flavor, the
helicity of a quark with $x\to 1$ must match the helicity of the parent
hadron, so that $\Delta q_{iV}(x)$ should approach $q_{iV}(x)$. For valence
$d$-quarks this means that $\Delta d_V(x)$, which is negative
at small and intermediate $x$\footnote{At these values of $x$, the
negative $\Delta d_V(x)$ was anticipated from theoretical
considerations, especially from the sign of its first moment
$\int^1_0\Delta d_V(x)dx \simeq -0.46 \pm 0.04$; these anticipations were
recently confirmed by the experimental values of $\Delta d_V(x)$, derived
from the analysis of nucleon and deuteron DIS data~\cite{Wislicki}.},
must change
sign at some $x$, typically chosen to be around $0.4\div 0.7$.

It is necessary to mention that equation (\ref{power}) provides only
upper limit for
the magnitudes of quark distributions, which is achieved only if
there are no other sources of suppression. Meanwhile, there is strong
experimental evidence that at large $x$ $d$-quarks are suppressed
compared to $u$-quarks.
In particular, most part of the experimental information about
the ratio of nucleon structure functions $F^n_2(x,Q^2)/F^p_2(x,Q^2)$ ,
including the most recent NMC data~\cite{NMC}, shows
that it falls lower
than the limiting value $3/7$ predicted by perturbative-QCD based approach.
The fit of the overall world data on $F^n_2/F^p_2$, presented in \cite{NMC},
tends to $1/4$, which is the limiting value for this ratio when $d_V(x)$
falls faster than $u_V(x)$.

Such a suppression of valence $d$-quarks is an essential feature of
another model, proposed by Carlitz and Kaur~\cite{CK}. The valence
distributions in this approach are built using simple quark
model considerations, especially the assumption that all diquarks
with $S=1,\, I=1$ are suppressed relative to those with $S=0$. The relation
$d_V(x)/u_V(x)\to1/5$, obtained in FJ model, should not hold in CK approach,
and valence quark distributions can be easily accommodated to satisfy the
tendency of $F^n_2/F^p_2$ to $1/4$. Just as in FJ model,
$\Delta u_V(x)/u_V(x)= 1$ as $x\to 1$, but the behavior of
valence $d$-quarks is different and $\Delta d_V(x)/d_V(x)=-1/3$.
For typical parameterizations of Carlitz-Kaur cosines, describing dilution of
the valence quark spin in the parton sea and quickly approaching $1$ at
intermediate and large $x$, $\Delta d_V/d_V$ is close to its limiting value
already at $x\sim0.3\div0.4$.

The latter distinction between FJ and CK models (the retention
of valence $d$-quark helicity in the first case and $d$-quark negative
polarization of $-1/3$ in the second) may serve as a basis for one more
experimental test of proton inner structure. The inclusive production of
$W^-$-bosons in $pp$- or $p\bar p$-collisions is very suitable for the
investigation of this distinction~\cite{Soffer}. Since parity is not
conserved in weak interactions, only one polarized initial beam is necessary
to get nonzero cross-section asymmetries. Thus, one opportunity for this
experiment may arise if a polarized proton beam will be put into operation at
TEVATRON.  It will be also possible to investigate $W^-$-boson asymmetries in
proton-proton interactions at BNL-RHIC collider.

On the parton level the leading contribution to $W^-$-production is
provided by the quark-antiquark annihilation $d\bar u \to W^{-}$.
The analysis of Born contribution at small and intermediate values of
$W^-$-boson rapidities $y$ has
already been carried out in~\cite{Soffer}; however, in this
paper I would like to discuss polarized $W^-$-production at large $y$,
where the cross-sections quickly fall down, and estimate possible errors
of the asymmetry measurement. Also, to get reliable estimates of
$W^{\pm}$-boson cross-sections, one has to consider higher-order
contributions; the discussion of them will be presented elsewhere~\cite{1st}.

For the reaction $p^{\ra}p\to W^- X$
the cross-section $d\sigma^{W^{-}}_{pp}/dy$ and the
corresponding single-spin asymmetry $A_L$ are written as~\cite{Soffer}
\begin{equation}
 \frac{d\sigma^{W^{-}}_{pp}}{dy}= \frac{1}{3}G_F\sqrt{2}\tau\pi
    \left( d(x_a,M^2_W)\, \bar u(x_b,M^2_W) + \bar u(x_a,M^2_W)
    \, d(x_b,M^2_W)\right) \label{cross}
\end{equation}
and
\begin{equation}
 A_L(y)=\frac{\Delta d(x_a,M^2_W)\, \bar u(x_b,M^2_W) - \Delta\bar
u(x_a,M^2_W) \, d(x_b,M^2_W)}{d(x_a,M^2_W)\, \bar u(x_b,M^2_W) + \bar
u(x_a,M^2_W)\, d(x_b,M^2_W)}.    \label{asym}
 \end{equation}
 Here $M_W$ and $y$
are the mass and rapidity of $W^{-}$-boson,
$G_F$ is the Fermi constant and
  \begin{displaymath} x_a=\sqrt{\tau}e^{y},\,
  x_b=\sqrt{\tau}e^{-y} \mbox{ with } \tau=M^2_W/s.
 \end{displaymath}

The $p\bar p$ cross-section $d\sigma^{W^{-}}_{p\bar p}/dy$ and asymmetry
$A_L$ can
be obtained from (\ref{cross},\ref{asym})
by substituting $\bar u(x_b,M^2_W)$
and $ d(x_b,M^2_W)$ correspondingly for $u(x_b,M^2_W)$ and $\bar
d(x_b,M^2_W)$.  It can be seen that when $y$ tends to upper kinematical
bound, $y\to -\ln \sqrt{\tau}$, $A_L$ is directly proportional to $\Delta
d(x_a,M^2_W)/d(x_a,M^2_W)$ at $x_a\to 1$.  In this limit the other Bjorken
variable, $x_b$ is of the order $M^2_W/s\sim 0.03$ for RHIC and 0.002 for
TEVATRON. For such $x_b$'s valence quarks become less important
compared to the sea, and therefore $u(x_b,M^2_W)\to \bar
  u(x_b,M^2_W)$, $d(x_b,M^2_W)\to \bar d(x_b,M^2_W)$. In its turn, this means
  that $pp$ and $p\bar p$ cross-sections, measured at same $\sqrt{s}$, at
large $y$ should be described by approximately the same dependence.  However,
as shown by numerical results, the cross-sections for typical RHIC
energies, assumed to be $\sqrt{s}=500\, GeV$, in the region of most interest
are smaller than TEVATRON ones, calculated for $\sqrt{s}=1.8\,
TeV$.

Fig.1 and 2 show unpolarized cross-sections $d\sigma^{W^{-}}_{pp}/dy$,
$d\sigma^{W^{-}}_{p\bar p}/dy$ and corresponding asymmetries $A_L$ as
functions of $x_a$, varying in the interval $0.4\leq x_a\leq 0.9$. These
values
of $x_a$ correspond to $W^-$-boson rapidities
$0.9\leq y\leq 1.7$ for RHIC and $2.2\leq y\leq 3$ for TEVATRON. The
cross-sections were calculated using
Gl\" uck-Reya-Vogt unpolarized distributions~\cite{GRV}.

\begin{figure}[t]
\epsfysize 3.5in
\epsffile{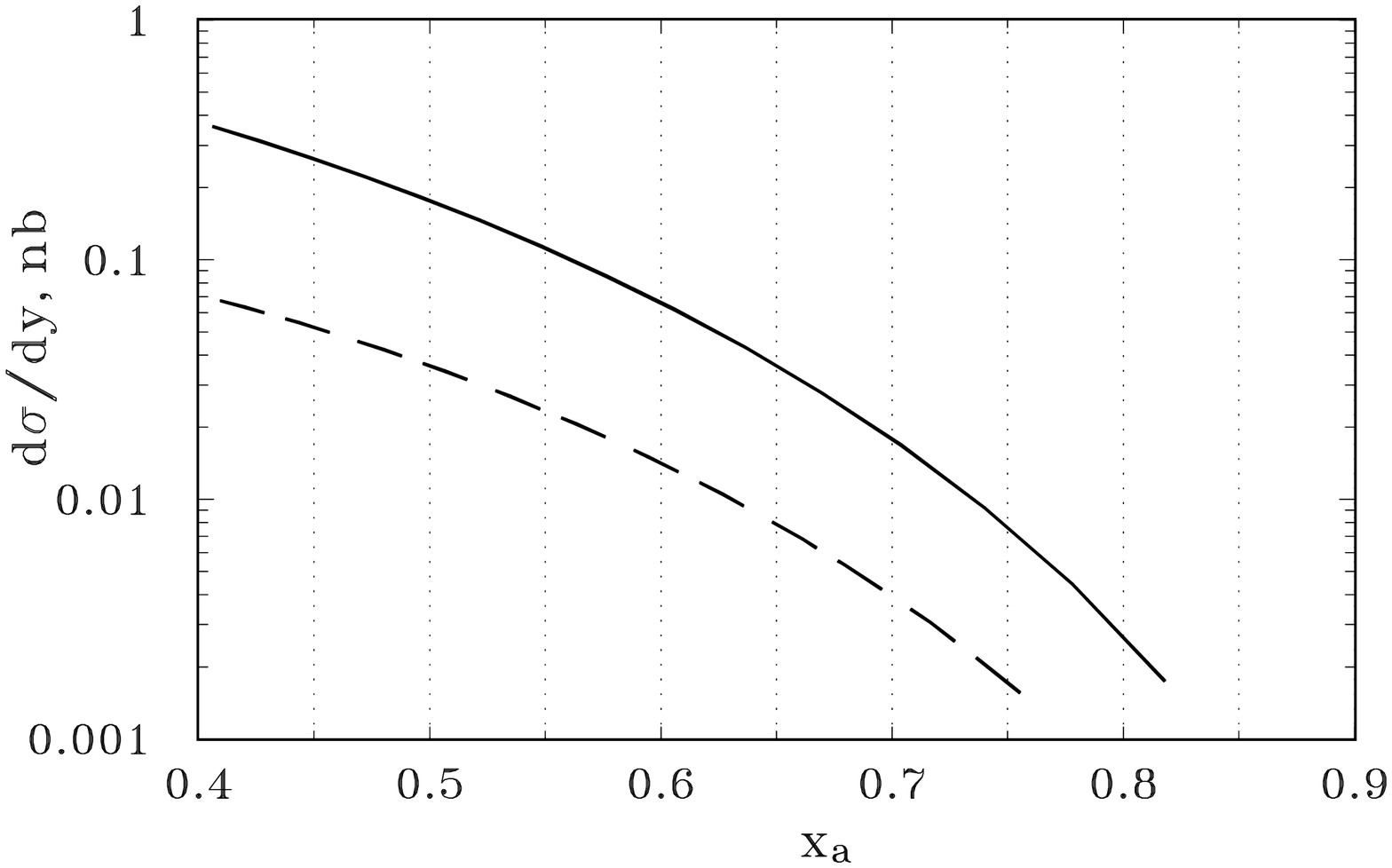}
{\small Fig.1. Differential cross-sections of inclusive $W^{-}$-boson
production in $pp$-collisions at RHIC ($\sqrt{s}=500\, GeV,$ dash line) and
in $p\bar p$-collisions at TEVATRON ($\sqrt{s}=1.8\, TeV,$ solid line).}
\end{figure}
\begin{figure}[ht]
\epsfysize 3.5in
\epsffile{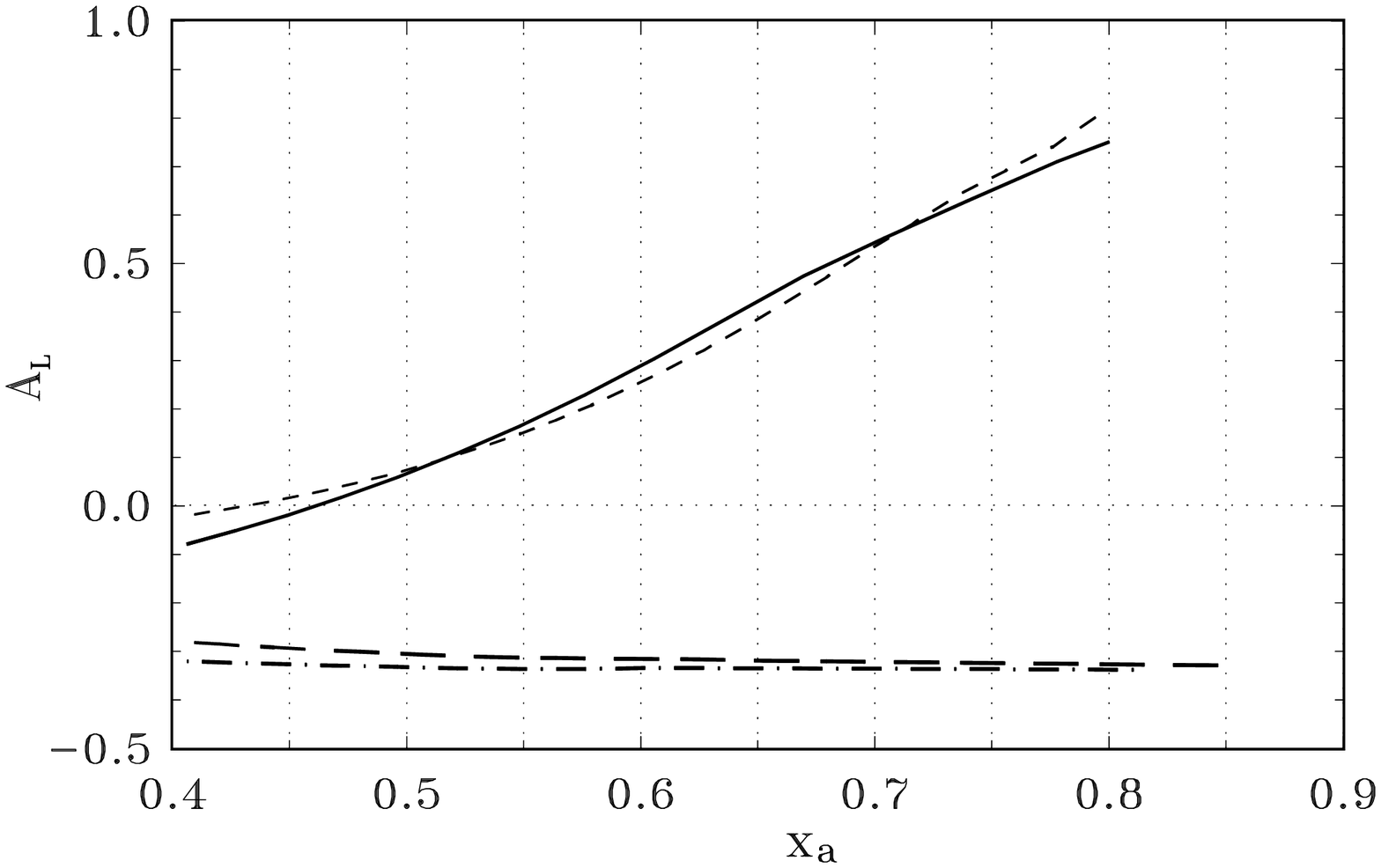}
{ \small Fig.2. Single-spin longitudinal asymmetries $A_L$ in
inclusive $W^{-}$-boson production. Short dash and solid lines: asymmetries
$A_L$ in $pp$- and $p\bar p$-channels for BBS model~\cite{BBS}.
Long dash and dot-dash lines: asymmetries $A_L$ in $pp$- and $p\bar
p$-channels for CK-type distributions~\cite{PN}}
\end{figure}

As can be seen from Fig.1, both $pp$- and $p\bar p$-cross-sections
quickly fall down with the increase of $x_a$,
$d\sigma^{W^{-}}_{pp}/dy$ being smaller  than $d\sigma^{W^{-}}_{p\bar
p}/dy$. Typical values of cross-sections at $x_a=0.4$ and $0.7$
are $0.07 $ and $0.003\, nbs$ in $pp$-channel, $0.3$ and $0.015\, nbs$
in $p\bar p$-channel. Knowing these cross-sections, it is
possible to
estimate the statistical error in determination of asymmetry $A_L$:
\begin{equation}
\delta A_L=\frac{1}{\it p_{beam}}\frac{1}{\sqrt{N}}=
\frac{1}{\it p_{beam}}\frac{1}{\sqrt{L\cdot T\cdot C}}\frac{1}{\sqrt{d\sigma
/dy}}.
\end{equation}
Here ${\it p_{beam}}$ is polarization of the beam and
$N$ is the number of events,
expressed as a product of the corresponding cross-section,
beam luminosity $L$, pure running time $T$ and combined trigger and
reconstruction efficiency $C$.
 Assuming that $L=10^{32}\,
cm^{-2}s^{-1}$ for RHIC and $10^{31}\, cm^{-2}s^{-1}$ for TEVATRON,
$T=10^7\, s$ (about 4 months), ${\it p_{beam}}=0.8$ and $C=0.5$
 it is easy to
obtain that
 \begin{equation}
 \delta A_L\Big\vert_{x_a=0.4}=0.007 ,\mbox{  } \delta
A_L\Big\vert_{x_a=0.7}=0.03 \label{errpp}
 \end{equation}
for RHIC and
 \begin{equation}
 \delta A_L\Big\vert_{x_a=0.4}=0.01 ,\mbox{  } \delta
A_L\Big\vert_{x_a=0.7}=0.045 \label{errpap}
\end{equation}
for TEVATRON.

Fig.2 shows the asymmetries $A_L$ in $pp$- and $p\bar p$-channels both for
CK and FJ-type polarized distributions, represented correspondingly by
set 1 distributions from~\cite{PN} and by recently proposed
Brodsky-Burkardt-Schmidt (BBS) distributions
from~\cite{BBS}\footnote{In BBS model $d_V(x)$ is not suppressed and
behaves
like $(1-x)^3$; this leads to larger unpolarized cross-sections of
$W^{-}$-production compared to those obtained with standard unpolarized
distributions. Nevertheless, the conclusions about
$A_L$ are influenced only slightly, since $A_L$ practically depends only on
the ratio $\Delta d_V(x)/d_V(x)$.}. As expected from the above discussion,
already at intermediate $x$ the behavior of $A_L$ in CK and FJ models is very
different: while the former is close to $-1/3$, the latter is around zero and
grows.  In BBS model $\Delta d_V(x)$ changes sign approximately at $x=0.5$.
It can be seen that even if the zero point is located at larger $x$, e.g.
$x=0.7$, the precision of the experiment still allows to reliably
discriminate one model from another.

To conclude, though the problem of the description of helicity-dependent
down quark distributions at large $x$ is very old and well-known, no
direct experimental test of $\Delta d_V(x)$ at $x\to 1$ has been done yet.
The study  of polarized $W^-$ production at RHIC or TEVATRON colliders
may provide important experimental information on the behavior of
$\Delta d_V(x)/d_V(x)$ and help clarify this interesting question.
\section*{Acknowledgements}
I express my thanks to S.M.\ Troshin for his help during the work on this
paper. I am also grateful to W.-D.\ Nowak, who provided me the
formula for estimate of the errors of asymmetry measurement.



\begin{thebibliography}{99}
\bibitem{EMCSMCE142}J.\ Ashman et al.: Nucl.Phys. B328 (1989) 1;
B.\ Adeva et al.: Phys.Lett. B302 (1993) 533;
D. Adams et al.: Phys. Lett. B329 (1994) 399;\\
P.\ Anthony et al.:  Phys.Rev.Lett 71 (1993) 959.
\bibitem{Wislicki}W.\ Wislicki, SMC Coll.: Talk presented at XXIX Rencontres
de Moriond, Meribel, France, March 1994 (to be published in Proceedings).
\bibitem{Close}F.E.\ Close: An Introduction to quarks and partons ---
Academic Press, London, 1979.
\bibitem{CK}R.D.\ Carlitz, J.\
Kaur: Phys.Rev.Lett. 38 (1977) 673;\\ J.\ Kaur: Nucl.Phys. B128 (1977) 219.
\bibitem{GunBrod}J.F.\ Gunion: Phys. Rev. D10 (1974) 242;
R.\ Blankenbecler and\\ S.J.\ Brodsky: Phys. Rev. D10 (1974) 2973.
\bibitem{FJ}G.R.\ Farrar, D.R.\ Jackson: Phys.Rev.Lett. 35
(1975) 1416.
\bibitem{NMC}P.\ Amaudruz et al.: Phys.Rev.Lett. 66 (1991)
2712.
\bibitem{Soffer}C.\ Bourelly, J.\ Soffer, F.M.\ Renard, P.\ Taxil:
Phys.Rep. 177 (1989) 319; C.\ Bourelly, J.\ Soffer: Nucl. Phys. B423 (1994)
329; preprint CPT-95/P.3160, February 1995 (to appear in Nucl.Phys. B).
\bibitem{1st}P.M.\ Nadolsky, in preparation.
\bibitem{GRV}M.\ Gl\" uck, E.\ Reya, A.\ Vogt: Z.Phys. C53 (1992) 127.
\bibitem{BBS}S.\ Brodsky, M.\ Burkardt, I.\ Schmidt: preprint SLAC-PUB-6087,
January 1994.
\bibitem{PN}P.\ Nadolsky: Z. Phys. C63, (1994), 601.
\end{thebibliography}
\end{document}